\documentclass[oribibl]{llncs}
\usepackage{hyperref}
\usepackage{graphicx}
\usepackage{color}
\usepackage{eurosym}
\usepackage[normalem]{ulem}

\title{Reasoning about Norms Revision}
\author{Davide Dell'Anna \and Mehdi Dastani \and Fabiano Dalpiaz}
\institute{Department of Information and Computing Sciences, Utrecht University,
	Princetonplein 5, 3584 CC Utrecht, The Netherlands\\
}

\begin{document}
	\maketitle
	\begin{abstract}
		Norms with sanctions have been widely employed as a mechanism for controlling and coordinating the behavior of agents without limiting their autonomy. The norms enforced in a multi-agent system can be revised in order to increase the likelihood that desirable system properties are fulfilled or that system performance is sufficiently high. In this paper, we provide a preliminary analysis of some types of norm revision: relaxation and strengthening. Furthermore, with the help of some illustrative scenarios, we show the usefulness of norm revision for better satisfying the overall system objectives.
	\end{abstract}
	
	\section{Introduction}
	Modern software systems execute in highly dynamic environments \cite{DBLP:journals/cacm/SommervilleCCKKKMP12}. Typical dynamic settings are open environments, where many (heterogeneous) autonomous software components coexist, interact and can join and leave as they please. 
	Dynamicity and uncertainty can be caused by many different (often unpredictable) factors: events from the operating environment, behaviors emerging from the effects of software executions and their interactions, the change of external norms \cite{DBLP:journals/ijbpim/RosemannRF08} and the impact of this on software behaviour, etc. \cite{DBLP:journals/re/WhittleSBCB10}.
	
	In order to guarantee desirable overall system level properties, in the multi-agent systems research field, norms with sanctions have been proposed as a means to control and coordinate the behavior of autonomous agents without limiting their autonomy \cite{DBLP:conf/kr/BoellaT04, DBLP:journals/ai/BullingD16}.
	
	
	In many cases, however, it is infeasible for a system designer to anticipate all the possible states that the software systems and its operating environment can reach during execution \cite{DBLP:journals/re/WhittleSBCB10}, and to define adequate norms
	for each of them. 
	Furthermore, system objectives are themselves in constant motion \cite{DBLP:journals/infsof/ZowghiG03, DBLP:books/sp/ess14/ErnstBJM14}: new goals arise while others are dropped, the desired qualities vary, and the relative priority of the objectives evolves.
	
	A static predefined set of norms may often result at runtime inadequate to guarantee the overall system objectives in various contexts \cite{DBLP:conf/caise/AliDGS11, DBLP:conf/sigsoft/LetierL04}.
	
	
	
	Dynamic update of norms at runtime is therefore one of the key factors to build a versatile normative multi-agent system, able to accommodate an heterogeneous population of agents and capable of ensuring the overall system objectives within a dynamic and uncertain environment \cite{DBLP:conf/ecai/KnobboutDM16}.
	
	In this paper we provide a preliminary analysis of some main types of \textit{norm revision} that can be applied to a running system with the goal of better satisfying the overall system objectives.
	
	We define the concept of norms revision as the act of replacing a set of norms with a new set. We specialize norms revision in three different types: relaxation, strengthening and generic alteration, which differ for the type of relationship that the revised set has with the original one.
	Finally, we provide a series of practical examples to clarify the different types of revisions and their possible applications.
	
	This paper should be considered as the initial step of a research on runtime supervision of normative systems. We do not report on technical nor experimental results, but rather lay down the foundations of our future research. 
	
	The remainder of this paper is structured as follows: Section \ref{sec:background} frames our work within the existing literature, Section \ref{sec:norms_revision} discusses the problem of norms revision, Section \ref{sec:examples} provides a series of examples of revisions of norms and possible applicative scenarios. Section \ref{sec:conclusion} ends the paper with concluding remarks and future work.

	\section{Background}
	\label{sec:background}
	Numerous papers over the years have focused on deciding and proving the correctness of normative systems by model checking formulas describing desired properties such as liveness or safety properties \cite{DBLP:conf/dagstuhl/DastaniGMT09, DBLP:conf/aamas/KnobboutD12, DBLP:conf/ijcai/AlechinaDL13}. Despite their elegance, these approaches do not fully cope properly with the dynamicity of today's complex systems, where the behavior of the system may change over time due to changes in the participating agents, their behaviors, or the active norms.
	
	As of today, there is no generally agreed formal methodology 
	to reason about the dynamics of norms and their impact on system specification. However some formal frameworks emerged in recent years to cope with these problems. In the rest of this section we discuss these related works on norm dynamics.
	
	Knobbout \textit{et al.} \cite{DBLP:conf/ecai/KnobboutDM16} propose a dynamic logic to formally characterize the dynamics of state-based and action-based norms. Both in Knobbout's works \cite{DBLP:conf/ecai/KnobboutDM16,DBLP:conf/jelia/KnobboutDM14} and in Alechina \textit{et al.}'s \cite{DBLP:conf/ijcai/AlechinaDL13}, norm change is intended as norm addition. Taking these approaches as our baseline, we aim to investigate further types of norms update in order to extend the existing framework for dynamic normative systems.

	Aucher \textit{et al.} \cite{DBLP:conf/lori/AucherGHL09} introduce a dynamic context logic in order to describe the operations of contraction and expansion of theories by introducing or removing new rules. Governatori \textit{et al.} \cite{DBLP:journals/igpl/GovernatoriR10} investigate from a legal point of view the application of theory revision to reason about legal abrogations and annulments. The types of revision presented in this paper can be related to theory revision, but we take a multi-agent systems standpoint, in which norms revision should be studied in terms of its impact on agents autonomy and we leave for future work the study of the impact of a revision on the existing normative system.
	
	Norms update has also been studied from the perspective of approximation \cite{DBLP:conf/atal/AlechinaDL14}, where an approximated version of a norm is formally obtained to cope with imperfect monitors for the original norm.
	The concept of approximation is similar to our notion of relaxation, however it is defined with respect to a specific monitor: an approximated norm is synthesized from the original one in order to maximize the violations detectable by the available imperfect monitor. 
	Here we assume perfect monitors and we focus instead on the relationship between different norms in terms of behaviors they allow or prohibit.
	
	Whittle and colleagues \cite{DBLP:journals/re/WhittleSBCB10} present early studies on relaxation of requirements of a software system. They define a requirements language for self-adaptive systems that allows to specify, with opportune operators, relaxed versions of a requirement during the requirement elicitation phase. While they focus on a language point of view, useful for the acquisition and specification of norms, in this paper we assume the norms are already specified and we provide an analysis of revision from a normative point of view based on a formal model of multi-agent system that abstracts from the language used to define the norms. 
	
	\section{Norms Revision}
	\label{sec:norms_revision}
	The class of models that we use to study and describe normative multi-agent systems are transition systems. These models consist of states and transitions between states. The idea is that such model describes all possible transitions of states that can occur within the system as a result of the actions that the agents perform. 
	
	\begin{definition}[Transition System]
		\label{def:transition_system}
		Given a set of atomic propositions $\textit{Props}$, a transition system is a tuple $M = (S, T, s_0, V)$ where $S$ is a set of states, $T \subseteq S \times S$ is a transition relation, $s_0 \in S$ is a distinguished initial state and $V$ is a valuation function $V: S \rightarrow \textit{Props}$ associating propositions with states (i.e., defining which propositions hold in a state).
	\end{definition}
	
	Given a transition system $M = (S, T, s_0, V)$, a \textit{path} $r$ through $M$ is a sequence $s_0, s_1, s_2, \ldots$ of states such that $s_iTs_{i+1}$ for $i=0,1,\ldots$. 
	
	Each path can be seen as a possible behavior of the system. We consider some of the behaviors as desired and others as undesired and we use the concept of \textit{norm} to identify these behaviors.
	
\begin{figure}[h]
		\centering
		\vspace{-20pt}
		\includegraphics[scale=0.7]{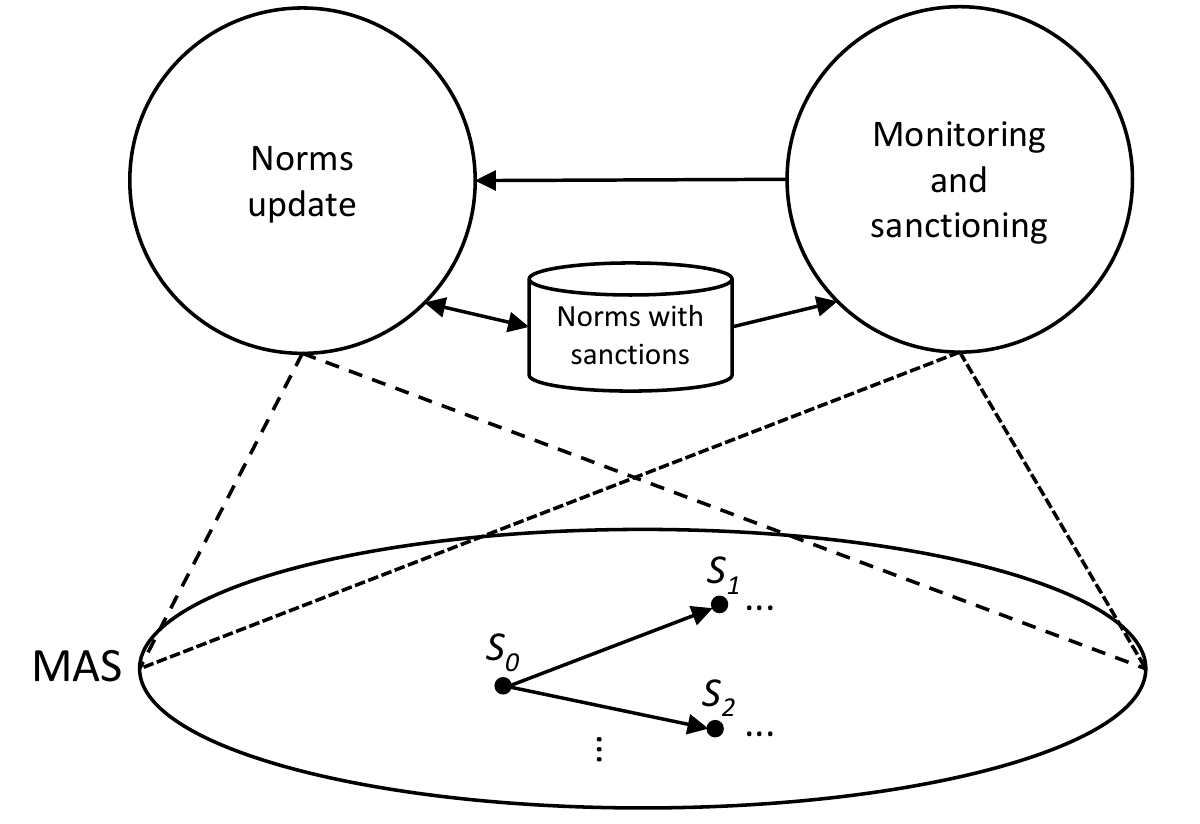}
		\vspace{-10px}
		\caption{The main components of the proposed runtime supervision framework 
			aimed at revising norms enforced into a normative MAS. Dotted lines represent the scope of the components.}
		\label{fig:architecture}
	\end{figure}

	In this paper we consider conditional norms with sanctions and deadlines, as they are commonly used in the existing literature on normative multi-agent systems \cite{DBLP:conf/iat/TinnemeierDMT09,DBLP:conf/ijcai/AlechinaDL13} and they have proven to be a reasonable compromise between expressiveness and ease of reasoning \cite{DBLP:conf/atal/AlechinaDL14}.
	\begin{definition}[Conditional Norms]
		\label{def:norms}
		Given a propositional language $L_N$, let $\textit{cond}$, $\phi$ and $d$ be boolean
		combinations of propositional variables from $L_N$ and let $\textit{san}$ be a propositional atom. A conditional norm $n$ is represented by a tuple
		$(\textit{cond};Z(\phi); d; \textit{san})$ where $Z$ can be either $O$ (obligation) or $F$ (prohibition). 
	\end{definition}
	Given a conditional norm, $cond$ represents the condition that must be satisfied in a state of $M$ in order to detach the norm. A detached norm persists as long as it is not obeyed or violated or the deadline $d$ is not reached (a state where $d$ holds is encountered).
	Conditional norms are evaluated on paths of the transition system. 
	In this paper we omit a formal definition of a violation, which is strictly dependent on the language used to express them.
	In the following we denote a violation formula for a conditional norm $n$ by $v(n)$ (see previous publication \cite{DBLP:conf/atal/AlechinaDL14} for a formal definition for PLTL formulae). A norm $n$ is violated on a path $r$ if, and only if, $v(n)$ holds in some state on $r$.  We denote by $\textit{Viol}(M, n)$ the set of paths through $M$ that violate the norm $n$. 
	Fig. \ref{fig:architecture} sketches the main components of a runtime supervision framework that continuously monitors the execution of a multi-agent system (\texttt{MAS}), evaluates its behavior against the currently enforced norms, and intervenes by deciding which norms should be revised.
	Norm violation is monitored (and sanctioned) by the \texttt{Monitoring and sanctioning} component. In this paper we ignore the techniques of monitoring and evaluation of norms, for which many works can be found in literature (e.g. \cite{DBLP:conf/atal/BullingDK13}), and we assume there exists a perfect monitor for each norm.
	The aim of this paper is to provide a study of the possible revisions of norms and of the possible conditions when a revision may be useful. The study is a starting point to build a \texttt{Norm update} component able to revise at runtime the currently enforced norms in order to ensure the achievement of the overall objectives. We assume that the \texttt{Monitoring and sanctioning} component stores at runtime information about the obedience or violation of the norms and about the operating contexts in which they are evaluated (e.g. the hour of the day, etc.). This information, together with information about the satisfaction of the overall systems objectives, is used by the \texttt{Norms update} component in order to decide if and how to revise the currently enforced norms.
	\vspace{-18px}
	
	Norm enforcement in multi-agent systems can be done in two possible ways: regimentation and sanctioning \cite{DBLP:conf/ijcai/AlechinaDL13}.
	Regimentation makes bad behaviors impossible, i.e., it makes some of the paths of $M$ inaccessible. With regimentation, violations of the norms are not possible, however the autonomy of the agents is reduced. Sanctioning norms violation is instead a means to discourage the violation. Sanctions are essentially treated like fines. They penalize agents when they bring the system on an undesired path, however they leave autonomy to the agents, even if their resources are reduced if they violate the norms.
	
	A regimented norm $n$ can be described by $\textit{Viol}(M, n)$ (i.e., by the set of paths of $M$ that violate the norm), while a norm $n$ with a sanction is described by both $\textit{Viol}(M, n)$ and a propositional sanction atom that is asserted in case of violation of $n$. 
	We assume in the following to be able to compare two sanctions (e.g., consider the case of numeric sanctions such as money).
	
%
	
	Given a set of norms $N = \{n_1, ..., n_k\}$, we denote by $\textit{Viol}(M, N)$ the set of all paths through $M$ each violating at least one norm in $N$ (i.e., $\textit{Viol}(M, N) = \textit{Viol}(M, n_1) \cup ... \cup \textit{Viol}(M, n_k)$), and by $\textit{San}$ a conjunction of all sanctions associated with the norms in $N$ (i.e., $\textit{San}_N = \textit{san}_1 \wedge ... \wedge \textit{san}_k$).
	
	Given a pair of sets of norms, $N_1$ and $N_2$, we denote by 
	\begin{itemize}
		\item $N_1 < N_2$: the fact that $\textit{Viol}(M, N_1) \supset \textit{Viol}(M, N_2)$. In this case, we say that $N_1$ is a set of norms more restrictive than $N_2$, or that $N_2$ is more relaxed than $N_1$.
		\item $N_1 \equiv N_2$: the fact that $\textit{Viol}(M, N_1) = \textit{Viol}(M, N_2)$. We say that the sets of norms $N_1$ and $N_2$ are equally restrictive.
	\end{itemize}

	Note that the relationships between norms above reported can only be satisfied by pairs of sets of norms such that $\textit{Viol}(M, N_1) \subseteq (\supseteq) \textit{Viol}(M, Nn_2)$.
	
	\begin{definition}[Norms Revision]
		\label{def:norms_revision}
		Let $N$ be a set of norms, a revision of $N$ is a replacement of $N$ with a new set $N^R$. $N^R$ may be a more/equivalently/less restrictive set of norms or an alternative set where no relationship can be stated.
	\end{definition}
	
	We distinguish two main types of revision: relaxation and strengthening, and we consider all the other types of revisions as regular alteration.
	
	\begin{definition}[Revision, Relaxation, Strengthening]
		\label{def:types_of_revision}
		A relaxation of a set of norms $N$ is a revision of $N$ with a new set $N^R$ such that $N^R > N$. A strengthening of a set of norms $N$ is a revision of $N$ with a new set $N^R$ such that $N^R < N$. Any other revision of $N$ with a new set $N^R$ such that either $N^R \equiv N$ or no relationship can be stated are regular alterations of $N$. 
	\end{definition}
	
	Given a pair of norms, $n_1$ and $n_2$ such that $n_1 = (\textit{cond}_1;Z(\phi_1);d_1;\textit{san}_1)$ and $n_2 = (\textit{cond}_2 ;Z(\phi_2); d_2; \textit{san}_2)$,
	\begin{enumerate}
		\item $n_2$ is a relaxation of $n_1$ if and only if at least one of the following holds (all else being equal):
		\begin{enumerate}
			\item \label{enum:relax_cond} $\textit{cond}_2$ is a stricter formula $\textit{cond}_1$.  Note that with stricter (less strict) formula we mean a formula that is satisfied in strictly less (more) paths of $M$. A stricter formula for a condition of a conditional norm makes therefore the norm applicable in fewer paths than the original.
			\item \label{enum:relax_formula} $\phi_2$ is a less strict formula than $\phi_1$ if $Z=O$ (obligation), or $\phi_2$ is a stricter formula than $\phi_1$ if $Z=F$ (prohibition). A less strict obligation (or a stricter prohibition) makes the norm violated in fewer paths, which means that more agents' behaviors are allowed.
			\item \label{enum:relax_deadline} $d_2$ is a less strict formula than $d_1$. Note that this is valid since we only accept propositional formulae, therefore a less strict deadline for a conditional norm means that it can be withdrawn in more states.
		\end{enumerate}
		\item $n_2$ is a strengthening of $n_1$ if and only if $n_1$ is a relaxation of $n_2$.
		\item $n_2$ is a regular alteration of $n_1$ if it is neither a relaxation nor a strengthening
	\end{enumerate}
	
	
	Note that while a revision of a single norm can be analyzed in terms of all its components, revision of a set of norms involving more than one norm can only be compared in terms of the set of paths of $M$ and the sanctions. A revision $N^R$ of a set $N$ obtained by relaxing some of the norms and strengthening some others can be compared to $N$ only in terms of the resulting set of behaviors allowed in $M$, since $\textit{Viol}(M, N^R)$ can be either a subset or a superset of $\textit{Viol}(M, N)$ or even a disjoint set. However if the revised norms are all relaxed (strengthened) the resulting revised set is a relaxed (strengthened) set of $N$.
	
	Notably, increasing or decreasing sanctions associated to norms, under the assumption that such sanctions are comparable, is an alternative means to influence the behavior of agents that goes beyond relaxation or strengthening of norms. Thus, increasing the sanctions $San_N$ associated with a set of norms $N$ is a way to bring the system to the desired paths defined by $N$, while decreasing the sanctions is a way to accept undesired paths. In the next section we provide an intuitive explanation of this concept with the help of illustrative examples.
	
	\section{Illustrative Examples}
	\label{sec:examples}
	In this section we provide some illustrative examples of the different types of revision of norms and some illustrative scenarios 
	where to revise norms may be useful to ensure the overall system objectives.
	Note that in this paper we are not interested in how the norms that we use are acquired.
	Moreover, to consider only relevant norms, we assume that we are provided with information about the rationale behind norms, that relates them to the overall system objectives. Finally, we assume every parameter of the norms as perfectly monitorable.
	
	
	\subsection{Relaxation}
	Consider norm $n_1$: ``\textit{cars speed in road $r$ shall be kept below $15km/h$, otherwise \euro$10k$ fine}", formally (for brevity assume car and road as implicit) $n_1 = (\textit{inRoad}, F(\textit{speedAbove15}), \top, \textit{10keuros} )$, where $\top$ provides ``always" interpretation.
	Possible examples of relaxations of $n_1$ are the following (possibly combined):
	\begin{itemize}
		\item $r_1 =  ((\textit{inRoad}\ \wedge\ \textit{trafficHigh}), F(\textit{speedAbove15}), \top, \textit{10keuros})$.\\
		$r_1$ differs from $n_1$ only for the condition: $(\textit{inRoad}\ \wedge\ \textit{trafficHigh})$ is a stricter formula than $\textit{inRoad}$. Since the condition is more specific and the rest of parameters is analogous (case \ref{enum:relax_cond} of Definition \ref{def:types_of_revision} holds), $r_1$ is applicable in fewer paths of $M$, therefore fewer paths are violations and more behaviors are allowed in the system.
		\item $r_2 =  (\textit{inRoad}, F(\textit{speedAbove20}), \top, \textit{10keuros})$.\\
		$r_2$ differs from $n_1$ only for the prohibition (case \ref{enum:relax_formula}): $\textit{speedAbove20}$ is stricter than $\textit{speedAbove15}$ because fewer paths are considered prohibited.
		\item $r_3 =  (\textit{inRoad}, F(\textit{speedAbove15}), \textit{firstHalfCompleted}, \textit{10keuros})$.\\
		$r_3$ differs from $n_1$ only for the deadline (case \ref{enum:relax_deadline}): $\textit{firstHalfCompleted}$ is less strict than $\top$ (``always", e.g. $\textit{firstHalfCompleted}\ \wedge\ \textit{secondHalfCompleted}$) since the norm remains valid only when the car is in the first half of the road. Again more behaviors are allowed. 
	\end{itemize} 
	
	\subsubsection{Example.} Consider a simple scenario where a population of autonomous vehicles can take a road $r$ where their speed and the traffic level can be perfectly monitored. The designer of the system $M$ wants to be sure that ``\textit{no cars stay in road $r$ ($10 km$ long) for more than 35 minutes}". Norm $n_1$ is currently enforced and at runtime it appears to be too strict for achieving the overall objective: if it is obeyed it prevents the fulfilment of the objective. 
	Since $n_1$ is too strict, one (or a combination) of the following relaxations may be useful to improve the performance of the system:
	\begin{itemize}
		\item replace $n_1$ with $r_1$, $r_2$ or $r_3$. In case of $r_1$, choosing a stricter condition allows agents to perform more actions without incurring in sanctions (agents are allowed to drive above 15km/h unless the traffic level is high). Since obeying to $n_1$ prevents the fulfillment of the objective, reducing the conditions where it applies means allowing behaviors previously forbidden. An analogous explanation can be provided also for $r_2$ and $r_3$.
	\end{itemize} 
	Notice that an analogous effect can be obtained by replacing $n_1$ with another norm $s_1 =  (\textit{inRoad}, F(\textit{speedAbove15}), \top, \textit{5euros})$, differing from $n_1$ only for the sanction: $\textit{5euros} < \textit{10keuros}$. Reducing the fine incentives more agents to violate $n_1$. Since the objective is achieved when $n_1$ is violated we increase the chances the objective is achieved.
	
	\subsection{Strengthening}
	Consider a norm $n_2 = (\textit{inRoad},\ O(\textit{speedbelow50}),\ \textit{outOfRoad},\ \textit{5euros})$.
	Possible examples of strengthening of $n_2$ are the following (possibly combined):
	\begin{itemize}
		\item $r_5 = ((\textit{inRoad}\ \vee\ \textit{aroundTheRoad}),\ O(\textit{speedbelow50}),$\\$(\textit{outOfRoad}\ \wedge\ \textit{1kmFarAway}),\ \textit{5euros})$.\\
		$r_5$ differs from $n_2$ for both the condition and the deadline: $(\textit{inRoad}\ \vee\ \textit{aroundTheRoad})$ is less strict than $\textit{inRoad}$, while $(\textit{outOfRoad}\ \wedge\ \textit{1kmFarAway})$ is stricter than \textit{outOfRoad}. Since the condition is less specific, $r_5$ is detached in more paths, and since the deadline is more specific, $r_5$ remains valid in more cases, therefore more paths violate it and less behaviors are allowed.
		\item $r_6 =  (\textit{inRoad}, O(\textit{typeScooter}), \textit{outOfRoad}, \textit{5euros})$.\\
		$O(\textit{typeScooter})$ is stricter than $O(\textit{speedbelow50})$:  while $n_2$ accepts all executions involving vehicles (scooters or not) that keep the speed below 50km/h, $r_6$ accepts only executions involving scooters (assumed here to have maximum speed of 50km/h). Again less behaviors are allowed.
	\end{itemize} 
	
	\subsubsection{Example.} Consider a scenario where a population of autonomous vehicles can take a road $r$ where their speed, type of vehicle and noise they produce can be perfectly monitored. There is an overall system objective ``\textit{vehicle's noise in the neighborhood is below $x$ db}". Norm $n_2$ is currently enforced and it is proven that violation of $n_2$ prevents the achievement of the overall objective. Norm $n_2$ appears at runtime to be too weak to ensure the overall system objective (e.g. it is often violated). 
	One (or a combination) of the following strengthening of $n_2$ may therefore be useful to improve the performance of the system:
	\begin{itemize}
		\item replace $n_2$ with $r_5$. Choosing a less strict condition makes the norm valid in more cases. Deincentivating agents to drive faster than 50km/h also around the road may prevent vehicles to accelerate immediately out of the road, which may be harmful for the achievement of the overall objective of keeping the noise in the neighborhood low.
		\item replace $n_2$ with $r_6$. Assuming scooters have maximum speed of 50km/h, if the new norm is obeyed by agents, the goal is ensured.
	\end{itemize} 
	
	Notice one more time that an analogous effect can be obtained by replacing $n_2$ with another norm $s_2 =  (\textit{inRoad}, O(\textit{speedbelow50}), \textit{outOfRoad}, \textit{10keuros} )$, differing from $n_2$ only for the sanction: $\textit{10keuros} > \textit{5euros}$. Increasing the fine incentives more agents to obey $n_2$. Since the objective is achieved when $n_2$ is obeyed we increase the chances to achieve it.
	
	\subsection{Alteration}
	Consider norm $n_3$: ``\textit{if two cars $c1$ and $c2$ are at the opposite ends of the narrowed road, $c1$ shall move and $c2$ shall wait, otherwise 10k euros fine to $c2$}", formally $n_3 = ((\textit{firstEnd}(\textit{c1})\ \wedge\ \textit{secEnd}(\textit{c2})),\ O(\textit{move}(\textit{c1})\ \wedge\ \textit{wait}(\textit{c2})),\ \top,\ \textit{c2\_10keuros})$.
	Possible examples of alteration of $n_3$ are the following:
	\begin{itemize}
		\item $r_8 = ((\textit{firstEnd}(\textit{c1})\ \wedge\ \textit{secEnd}(\textit{c2})),\ O(\textit{wait}(\textit{c1})\ \wedge\ \textit{move}(\textit{c2})),\ \top,\ \textit{c1\_10keuros})$.
		\item $r9 = (\textit{inRoad}(c1),\ O(\textit{speedbelow50}),\ \top,\ \textit{c1\_10keuros})$.
	\end{itemize}
	Both $r8$ and $r9$ define a set of underused paths that is neither a subset or a superset of the paths defined by $n_3$. Therefore they cannot be considered neither a relaxation nor a strengthening. Notice that $r9$ is reported as an example of a generic alteration, which reflects the definition we provided, even if it is not strictly related to $n_3$.
	\subsubsection{Example.} Consider a scenario of a narrowed road and cars coming from both directions. The overall objective is ``\textit{keep cars queue size in both directions below a threshold t}". Norm $n_3$ is currently enforced and it proved to be useful to achieve the objective when traffic is low in both directions. However, when traffic is high in the direction of car $c1$ the queue in the opposite direction grows too much. Norm $n_3$ appears therefore to be not good to ensure the overall system objective in such case.
	An alteration such as replacing $n3$ with $r9$ may be useful to improve the performance of the system.
	Notice that a generic alteration involves a new norm that either is not strictly related to the existing one or it is equally restrictive, therefore it is hard to predict the outcome of the enforcement of the new norm. However, assuming the considered norms are related to the same overall system objective, enforcing an alteration in the system may lead to different (hopefully better) performances.
	
	\section{Conclusion and Future Work}
	\label{sec:conclusion}
	In this paper we proposed a preliminary study of some types of revision of norms that can be enacted on a running multi-agent system in order to enhance its performance.
	We defined the concept of norms revision as a replacement of a set of norms with a new set and we formally defined the notions of relaxation and strengthening, which introduce supersets and subsets of the allowed behaviors, respectively. We provided some examples of revisions and some application scenarios where a revision of the current set of enforced norms may be useful in order to achieve the overall system objectives.
	
	The analysis of different types of norms update provided in this paper is a first step in order to extend existing works for dynamic normative systems (see \cite{DBLP:conf/jelia/KnobboutDM14, DBLP:conf/ecai/KnobboutDM16}) with revision of norms.
	As future work we plan to provide a formal semantics for the concepts here introduced and to develop a syntactically sound and complete reasoning system. We also want to consider the effects of a revision in terms of update of the normative system and to analyse the relation of the revisions here presented with standard operators from the literature (e.g. from AGM framework).
	Finally we plan to provide a formal discussion about the correlation between the enforced norms and the fulfilment of the overall system objectives and to develop techniques to automatically reason at runtime about this correlation and to automatically suggest and perform a revision. These developments are meant to be part of an adaptive runtime supervision framework that continuously monitors the execution of a normative multi-agent system and intervenes on it in order to improve its performance.
	
	\bibliographystyle{splncs}
	\bibliography{Bibliography}
\end{document}